\documentclass[runningheads]{llncs}
\usepackage{graphicx}
\usepackage{xspace}
\usepackage{amssymb}
\usepackage{amsmath}
\usepackage{makecell}
\usepackage{booktabs} % For formal tables
\usepackage{todonotes}
\usepackage{graphicx}
\usepackage{epstopdf}
\usepackage{url}
\usepackage{algorithm}
\usepackage[noend]{algpseudocode}
\usepackage{soul}
\usepackage{graphicx}
\usepackage{subfig}
\usepackage{multirow}
\usepackage{wrapfig}
\usepackage{setspace}
\def\framework{\texttt{TimePop}\xspace}
\def\TopN{\textit{top-N}\xspace}

\begin{document}
\title{Local Popularity and Time in top-N Recommendation}
%\author{ }
%\institute{}
%\author[V.W. Anelli et al.]{Vito Walter Anelli$^\star$, Tommaso Di Noia$^\star$, \\Eugenio Di Sciascio$^\star$, Azzurra Ragone$^\bullet$, Joseph Trotta$^\star$}
%\institute{%
%	$^\star$Polytechnic University of Bari\\
%	%  \streetaddress{P.O. Box 1212}
%	Bari - Italy\\
%	{firstname.lastname}@poliba.it\\
%	$^\bullet$Independent Researcher\\
%	azzurra.ragone@gmail.com
%	%  \postcode{43017-6221}

%\author{Vito Walter Anelli\inst{1}\orcidID{0000-1111-2222-3333} \and
%Second Author\inst{2,3}\orcidID{1111-2222-3333-4444} \and
%Third Author\inst{3}\orcidID{2222--3333-4444-5555}}
%% First names are abbreviated in the running head.
%% If there are more than two authors, 'et al.' is used.
%%
%\institute{Princeton University, Princeton NJ 08544, USA \and
%Springer Heidelberg, Tiergartenstr. 17, 69121 Heidelberg, Germany

\author{Vito Walter Anelli\inst{1} \and
	Tommaso Di Noia\inst{1} \and
	Eugenio Di Sciascio\inst{1} \and
	Azzurra Ragone\inst{2} \and
	Joseph Trotta\inst{1}
}
\authorrunning{V.W. Anelli et al.}
\institute{Polytechnic University of Bari, Bari, Italy \\
	\email{\{firstname.lastname\}@poliba.it} \and
	Independent Researcher \\
	\email{azzurra.ragone@gmail.com}}
\maketitle              % typeset the header of the contribution
\begin{abstract}
Items popularity is a strong signal in recommendation algorithms. It strongly affects collaborative filtering approaches and it has been proven to be a very good baseline in terms of results accuracy. Even though we miss an actual personalization, global popularity can be effectively used to recommend items to users. In this paper we introduce the idea of a \textit{time-aware personalized popularity} in recommender systems by considering both items popularity among neighbors and how it changes over time. An experimental evaluation shows a highly competitive behavior of the proposed approach, compared to state of the art model-based collaborative approaches, in terms of results accuracy.

%\keywords{First keyword  \and Second keyword \and Another keyword.}
\end{abstract}
\section{Introduction}
\setstretch{0.93}
%\vspace{-0.3cm}
Collaborative-Filtering (CF) \cite{rendle2010factorization} algorithms, more than others, have gained a key-role among recommendation approaches and have been effectively implemented in commercial systems to help users in dealing with the information overload problem. Some of them also use additional information (hybrid approaches) to build a more precise user profile in order to serve a much more personalized list of items \cite{anelli2017feature,fernandez2016alleviating}.

However, it is well known \cite{JannachLGB13} that all the algorithms based on a CF approach are affected by the so called ``popularity bias'' meaning that  popular items tend to be recommended more frequently than those in the long tail. 
%The main assumption behind a ``most popular'' approach is that global popularity is a characteristic influencing all the users. 
Initially considered as a shortcoming of collaborative filtering algorithms and then not useful to produce good recommendations \cite{DBLP:conf/recsys/JamborW10}, in some works items popularity has been intentionally penalized \cite{DBLP:conf/icdm/OhPYSP11}. Very interestingly, a recommendation algorithm purely based on most popular items,  has been proven to be a strong baseline \cite{Cremonesi2010} although it does not exploit any actual personalization. More recently, popularity has been also considered as a natural aspect of recommendation that, by measuring the user tendency to diversification, can be exploited to balance the recommender optimization goals \cite{DBLP:journals/eswa/JugovacJL17}. The study of popularity in user tendencies is not completely new in the recommender systems field. Some interesting works explored these criteria for re-ranking purposes \cite{DBLP:journals/eswa/JugovacJL17,DBLP:conf/icdm/OhPYSP11}, and multiple goals optimization \cite{DBLP:conf/recsys/JamborW10}.

In the approach we present here, we introduce a more fine-grained personalized version of popularity by assuming that it is conditioned by the items that a user $u$ already experienced \textit{in the past}. To this extent, we look at a specific class of neighbors, that we name \textit{Precursors}, defined as the users who already rated  the same items of $u$ in the past. This led us to the introduction of a time-aware analysis while computing a recommendation list for $u$. 
As time is considered a contextual feature, most of the works dealing with temporal aspects are considered as a sub-class of Context-Aware RS (CARS) \cite{adomavicius2011context}: Time-Aware RS (TARS) \cite{zimdars2001using,adomavicius2001multidimensional,koren2010collaborative}. 
%Moreover it is already proven that using contextual informations \cite{adomavicius2011context,baltrunas2014experimental} (and time \cite{koren2010collaborative}) can lead to achieve better performance. 
In TARS, the freshness of  different ratings is often considered as a discriminative factor between candidate items. 
Usually, a time window \cite{lathia2009temporal} is adopted to filter out all the ratings that stand before (and/or after) a certain time relative to the user or the item. Recently, an interesting work that makes use of time windows has been proposed in \cite{DBLP:conf/recsys/BelloginS17} where the authors focus on the last common interaction between the target user and her neighbors to populate the candidate items list. 
In \cite{DBLP:journals/dss/BaoLLSG13} social information and time are integrated dealing with the interests of the users as a series of temporal matrices. Probabilistic matrix factorization technique are adopted to learn latent factors. Regarding sequences and recommendation it is worth to mention \cite{DBLP:conf/wsdm/WuABSJ17}, in which the authors combine an LSTM network with a low-rank matrix factorization algorithm to produce recommendation lists.
A pioneer work was proposed more than a decade ago in \cite{ding2005time} which used an exponential decay function $e^{-\lambda t}$ to penalize old ratings. An exponential decay function \cite{koren2010collaborative} was then used to integrate time in a latent factors model.
In the last years, several Item-kNN \cite{liu2010online,ding2005time} with a temporal decay function have been deployed. 
%Similarly a User-kNN model could be enhanced with a temporal decay function, this variant is also used in \cite{DBLP:conf/recsys/BelloginS17}. 
%Moreover we consider \cite{koren2010collaborative} relevant because of the $\Delta T$ used, as we did in our work. 
Another interesting work was proposed in \cite{xia2010dynamic} where three different kinds of time decay were adopted: exploiting concave, convex and linear functions. 

In this paper we present \framework, an algorithm that combines the notion of personalized popularity conditioned to the behavior of users' neighbors while taking into account the temporal dimension. 
It is worth noticing that \framework works with implicit feedback to compute recommendations.
%These contributions are weighted with an exponential decay function, with a few modifications (see Section \ref{sec:decay}) to make the approach more flexible and precise independently of the adopted dataset. 
Differently from some of the approaches previously described, in \framework we avoid both the use of a time window and the selection of a fixed number of candidate items. Indeed, while on the one hand, a time window may severely restrict the selection of candidates, on the other hand, a  fixed number of candidate items may heavily affect the algorithm results.

\section{Time-aware Local Popularity}\label{sec:approach}
%\vspace{-0.3cm}
%In this section we describe the idea and the technical details behind \framework, a time-aware popularity-based recommendation algorithm. 
The leading intuition behind \framework is that the popularity of an item has not to be considered as a global property but it can be personalized if we consider the popularity in a neighborhood of users.
% meaning that there is a degree of popularity that depends on the user and their neighbors. 
% As a matter of fact, a recommendation list based on a global popularity is exactly the same for all the users with no personalization. 
We started from this observation to formulate a form of personalized popularity, and then we added the temporal dimension to strengthen this idea.  %%%%% FIN QUI %%%%

In \framework, given a user $u$ the first step is then the identification of user's neighbors who rated the same items as $u$ but before $u$. We name these users \textit{Precursors}. In our intuition, Precursors represent a community of users $u$ relies on to choose the items to enjoy.
% if the popularity of an item is a personalized dimension we can  Hence, the first step is to identify and formally define these subgroups of users. 
%In general, we may say that users are supposed to process and exploit information (such as popularity) analogously to people that share with them tastes or way of thinking. 
In a neighborhood of $u$,  the same item is enjoyed by users in different time frames. 
%In our model, among these users,  we consider those who already enjoyed the same items of $u$ but before she did it. 
This leads us to the second ingredient behind \framework: personalized popularity is a function of time. The more the ratings about an item are recent, the more its popularity is relevant for the specific user. Hence, in order to exploit the temporal aspect of these ratings, the contributions of \textit{Precursors} can be weighted depending on their freshness.

We now introduce some basic notation that will be used in the following. We use $u \in U$ and $i \in I$ to  denote users and items respectively. Since we are not just interested in the items a user rated but also at when the rating happened, we have that for a user $u$ the corresponding user profile is $P_u = \{(i_1,t_{ui_1}), \ldots, (i_n,t_{ui_n})\}$ with $P_u \subseteq I \times \Re$, being $t_{ui}$ a timestamp representing when $u$ rated $i$.  
\begin{definition}[Candidate Precursor and Precursor]\label{def:candidate-precursor}
	%Let  $u \in U$ represent a set of users  and $i \in I$ a set of items. We use the pair $\langle u,(i, t_{ui})\rangle$ with $t_{ui} \in \Re$, to say that $u$ rated $i$ at time $t_{ui}$. 
	Given  $(i, t_{ui})\in P_u$ and $(i, t_{u'i})\in P_{u'}$, we say that $u'$ is a  \textbf{Candidate Precursor} of $u$ if $t_{u'i} < t_{ui}$. We use the set $\hat{\P}^u$ to denote the set of Candidate Precursors of $u$.
	Given two users $u'$ and $u$ such that $u'$ is a Candidate Precursor of $u$ and a value $\tau_u \in \Re$ we say that $u'$ is a \textbf{Precursor} of $u$ if the following condition holds.
	\[
	\lvert\{i \mid (i,t_{ui}) \in P_u \wedge (i,t_{u'i}) \in P_{u'} \wedge t_{u'i} < t_{ui})\}\rvert \geq \tau_u
	\]
	We use $\P^u$ to denote the set of \textbf{Precursors} of $u$.
\end{definition}
A user $u'$ is a \textit{Candidate Precursor} of $u$ if $u'$ rated at least one common item $i$ before $u$. Although this definition catches the intuition behind the idea of \textit{Precursors}, it is a bit weak as it considers also users $u'$ who have only a few or even just one item in common with $u$ and rated them before she did. Hence, we introduced a threshold taking somehow into account the number of common items in order to enforce the notion of \textit{Precursors}. 
%This threshold can be personalized or computed automatically (see Equation (\ref{eqn:threshold})).
The threshold parameter $\tau_u$ in Defintion \ref{def:candidate-precursor} can be also computed automatically as:
\begin{equation}\label{eqn:threshold}
	\tau_u = \frac{\sum_{u' \in \hat{\P}^u} \lvert\{i \mid (i,t_{ui}) \in P_u \wedge (i,t_{u'i}) \in P_{u'} \wedge t_{u'i} < t_{ui})\}\rvert}{|\hat{\P}^u|}
\end{equation}\vspace{-0.3cm}

\begin{wrapfigure}{r}{0.5\textwidth} 
	\vspace{-1.5cm}
	\begin{center}
		\centering
		\includegraphics[width=0.8\linewidth]{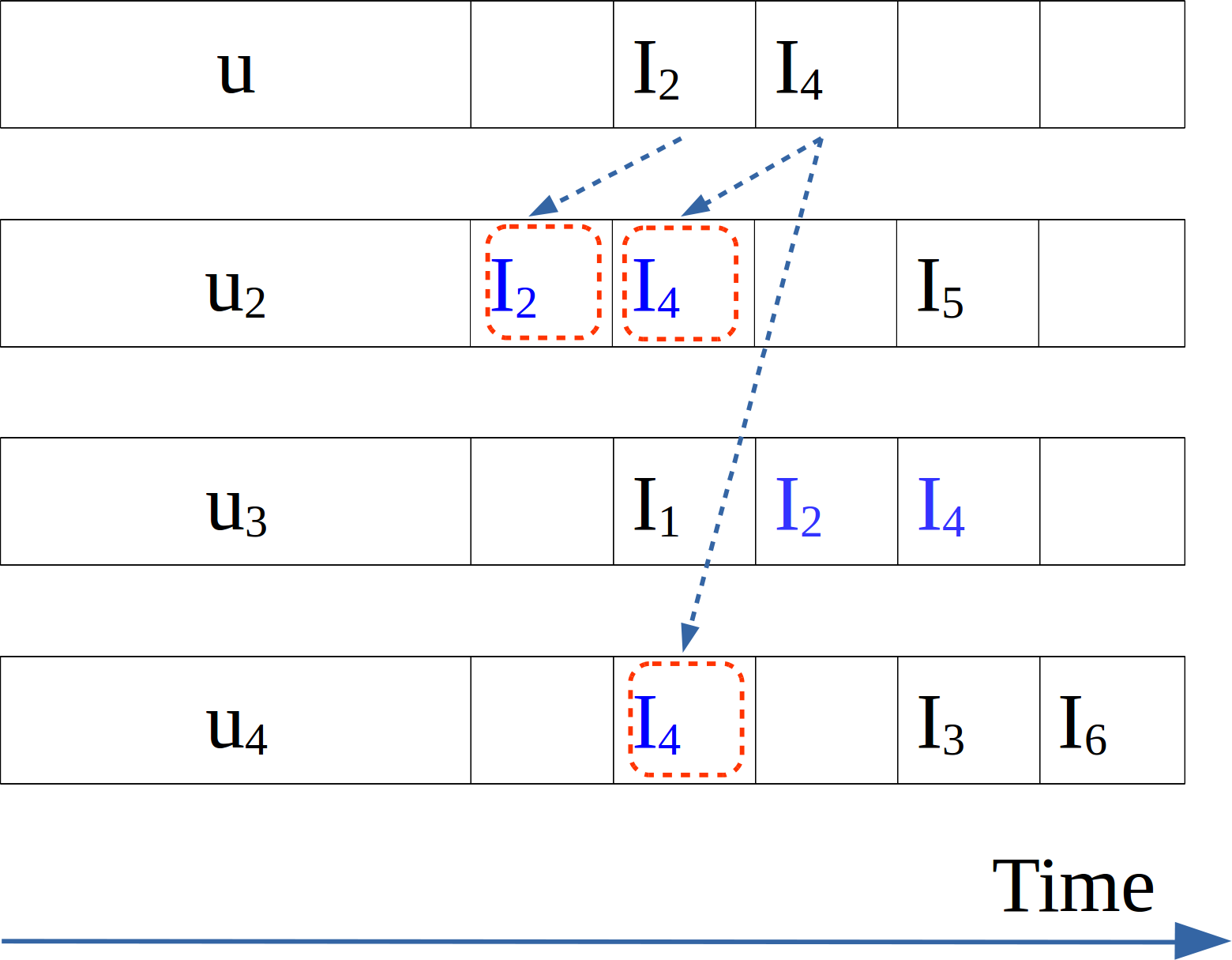}\vspace{-0.1cm}
		\caption{Example of Precursors computation.}
		\vspace{-1.1cm}
		\label{fig:Precursors}
	\end{center}
\end{wrapfigure}

\noindent To give an intuition on the computation of Precursors and of $\tau_u$ let us describe the simple example shown in Figure \ref{fig:Precursors}.\\
Here, for the sake of simplicity, we suppose that there are only four users and six items and $u$ is the user we want to provide recommendations to. Items that users share with $u$ are highlighted in blue  and items with a dashed red square are the ones that have been rated before $u$. We see that $\hat{\P}^u = \{u_2,u_4\}$. Indeed, although $u_3$ rated some of the items also rated by $u$ they have been rated after. By Equation (\ref{eqn:threshold}) we have $\tau_u = \frac{3}{2} = 1.5$. Then, only $u_{2}$ results to be in $\P^u$ because she has $2 > 1.5$ shared items rated before those of $u$. As for $u_3$, it is more likely that $u$ is a Precursor of $u_3$ and not vice versa.\vspace{-0.2cm}
\paragraph{Temporal decay.}\label{sec:decay}
As the definition of Precursor goes through a temporal analysis of user behaviors, we may look at the timestamp of the last rating provided by a Precursor in order to identify how active she is in the system. Intuitively, the contribution  to popularity for users who have not contributed recently with a rating is lower than ``active'' users.  On the other side, given an item in the profile of a Precursor we are interested in the freshness of its rating. As a matter of fact, old ratings should affect the popularity of an item less than newer ratings. Summing up, we may classify the two temporal dimensions as \textbf{old/recent user} and \textbf{old/recent item}. In order to quantify these two dimensions for Precursors we introduce the following timestamps: 
$\mathbf{t_0}$ this is the reference timestamp. It represents the ``now'' in our system; $\mathbf{t_{u'i}}$ is the time when $u'$ rated $i$; $\mathbf{t_{u'l}}$ represents the timestamp associated to the last item $l$ rated by the user $u'$.
Different temporal variables are typically used \cite{ding2005time,koren2010collaborative}, and they mainly focus on \textbf{old/recent items}. $\Delta T$ may refer to the timestamp of the items with reference to the last rating of $u'$ \cite{ding2005time} with  $\Delta T = \mathbf{t_{u'l}} - \mathbf{t_{u'i}}$ or to the reference timestamp \cite{koren2010collaborative} with  $\Delta T = \mathbf{t_0} - \mathbf{t_{u'i}}$.
As we stated before, our approach captures the temporal behavior of both \textbf{old/recent users} and \textbf{old/recent items} at the same time. We may analyze the desired ideal behavior of $\Delta T$ depending on the three timestamps previously introduced as represented in Table \ref{tbl:decay}.\vspace{-0.4cm}
\begin{wraptable}{l}{0.45\linewidth}\vspace{-0.2cm}
	\begin{tabular}{|l|l|l|}
		\hline
		& \thead{\textbf{recent user} \\ ($\mathbf{t_0} $ $\approx$ $\mathbf{t_{u'l}}$)} & \thead{\textbf{old user}\\($\mathbf{t_0} $ $\gg$ $\mathbf{t_{u'l}}$) } \\ \hline
		\thead{\textbf{recent item} \\($\mathbf{t_{u'l}}$ $\approx$ $\mathbf{t_{u'i}}$)} & 
		\makecell{$\approx0$} & 
		\makecell{$\mathbf{t_0}  - \mathbf{t_{u'l}}$} \\ \hline
		\thead{\textbf{old item}\\($\mathbf{t_{u'l}}$ $\gg$ $\mathbf{t_{u'i}}$)} & 
		\makecell{$\mathbf{t_{u'l}} - \mathbf{t_{u'i}}$} & 
		\makecell{$\mathbf{t_0}  - \mathbf{t_{u'l}}$} \\ \hline
	\end{tabular}\caption{\textit{Ideal values of $\Delta T$ w.r.t. the Precursor characteristics}}\vspace{-0.45cm}\label{tbl:decay}
	\vspace{-0.7cm}
\end{wraptable}

Let us focus on each case. In the upper-left case we want $\Delta T$ to be as small as possible because both $u'$ and the rating for $i$ are ``recent'' and then highly representative for a popularity dimension. In the upper-right case, the rating is recent but the user is old. The last item has been rated very close to $i$ but a large value of $\Delta T$ should remain because the age of $u'$ penalizes the contribution. The lower-left case denotes a user that is active on the system but rated $i$ a long time ago. In this case the contribution of this item is almost equal to the age of its rating. The lower-right case is related to a scenario in which both the rating and $u'$ are old. In this scenario, the differences between the reference timestamp minus the last interaction and the reference timestamp minus the rating of $i$ are comparable: $(\mathbf{t_0}  - \mathbf{t_{u'l}}) \approx (\mathbf{t_0}  - \mathbf{t_{u'i}})$. 
%We may assume that the last interaction and the rating are approximately equal because the difference between them is not relevant with respect to the other differences we mentioned. 
In this case, we wish the contribution of $\Delta T$  to consider the elapsed time from the last interaction (or the rating) until the reference timestamp. All the above observations lead us to define $\Delta T = \lvert \mathbf{t_0}  - 2\mathbf{t_{u'l}} + \mathbf{t_{u'i}} \rvert$.
In order to avoid different decay coefficients, in our experimental evaluation, all $\Delta Ts$ are transformed in days (from milliseconds) as a common practice.
\\\\\textbf{The  Recommendation Algorithm.}
We modeled our algorithm \framework to solve a \TopN recommendation problem. Given a user $u$, \framework computes the recommendation list by executing the following steps:
\\ \textbf{1.} Compute $\P^u$;
\\ \textbf{2.} For each item $i$ such that there exists $u' \in \P^u$ with $(i,t_{u'i}) \in P_{u'}$ compute a score for $i$ by summing the number of times it appears in $P_{u'}$ multiplied by the corresponding decay function;
\\ \textbf{3.} Sort the  list in decreasing order with respect to the score of each $i$.
\\For sake of completeness, in case there were no precursors for a certain user, a recommendation list based on global popularity is returned to $u$.
Moreover, if \framework is able to compute only $m$ scores, with $m < N$,  the remaining items are returned based on their value of global popularity.

\section{Experimental Evaluation}\label{sec:eval}
\vspace{-0.3cm}
In order to evaluate \framework we tested our approach considering datasets related to different domains. Two of them related to the movie domain ---the well-known Movielens1M dataset  and Amazon\footnote{\url{http://jmcauley.ucsd.edu/data/amazon/}} Movies --- and a dataset referring to toys and games ---Amazon Toys and Games, with 2M ratings and a sparsity of 99.99949\%. 
``All Unrated Items'' \cite{steck2013evaluation} protocol has been chosen to compare different algorithms where, for each user, all the items that have not yet been rated by the user all over the platform are considered.  In order to evaluate time-aware recommender systems in an offline experimental setting, a typical k-folds or hold-out splitting would be ineffective and unrealistic. 
To be as close as possible to an online real scenario we used the fixed-timestamp splitting method  \cite{DBLP:journals/umuai/CamposDC14,DBLP:reference/sp/GunawardanaS15}, also used in \cite{DBLP:conf/recsys/BelloginS17} but with a \textit{dataset centered} base set. 
The basic idea is choosing a single timestamp that represents the moment in which test users are on the platform waiting for recommendations. Their past corresponds to the training set, and the performance is evaluated with data coming from their future. In this work, we select the splitting timestamp that maximizes the number of users involved in the evaluation by setting two constraints: the training set must keep at least 15 ratings, and the test set must contain at least 5 ratings.
Training set and test set for the three datasets are publicly available\footnote{\url{https://github.com/sisinflab/DatasetsSplits}} along with the splitting code for research purposes.
In order to evaluate the algorithms we measured \textit{normalized Discount Cumulative Gain}@N ($nDCG@N$) using \textit{Time-independent rating order condition \cite{DBLP:journals/umuai/CamposDC14}}. The metric was computed per user and then the overall mean was returned using the RankSys framework and adopting \textit{Threshold-based relevant items} condition \cite{DBLP:journals/umuai/CamposDC14}. 
The threshold used to consider a test item as relevant has been set to the value of 4 w.r.t. a 1-5 scale for all the three datasets.
\vspace{-1cm}

\begin{figure*}[ht]
	\centering
	\begin{tabular}{ccc}
		%		\multirow{-3}{*}{\subfloat[legend]{\includegraphics[width=2cm]{imgs/new_images/legend_small.png}}}\\
		\subfloat[AmazonMovies]{\includegraphics[width=0.315\textwidth]{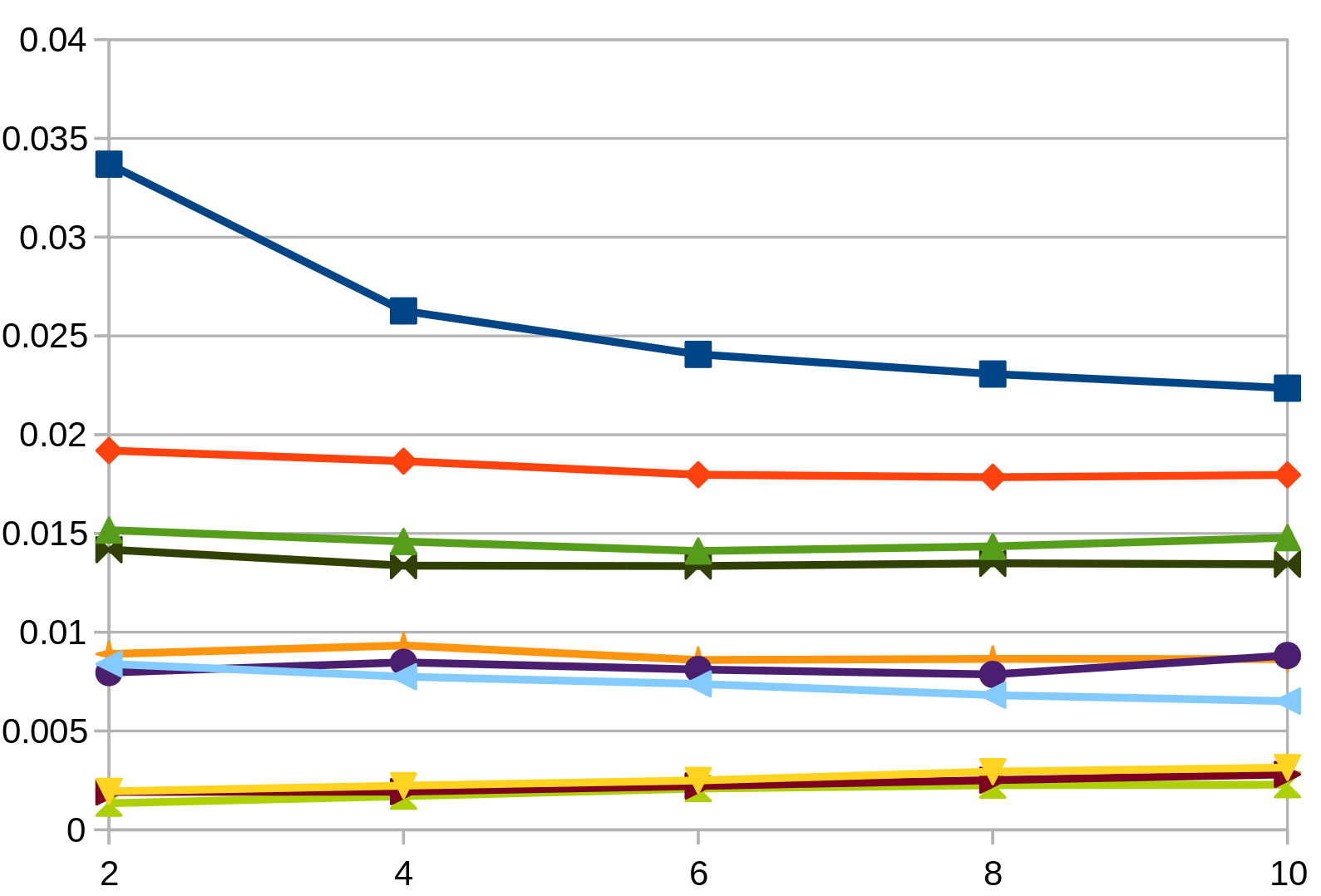}}  & 
		\subfloat[Movielens1M]{\includegraphics[width=0.315\textwidth]{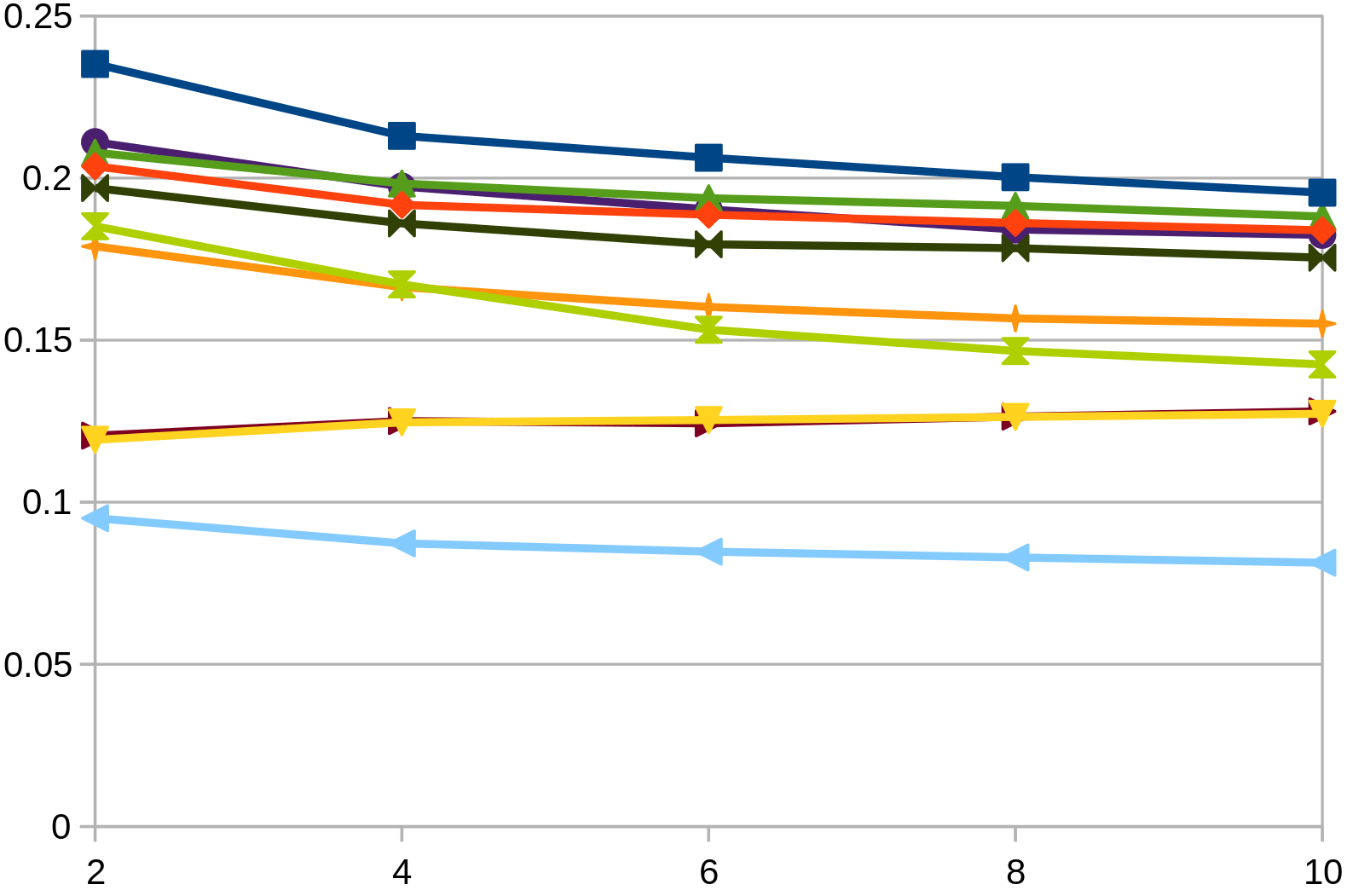}} & 
		\subfloat[AmazonToys]{\includegraphics[width=0.315\textwidth]{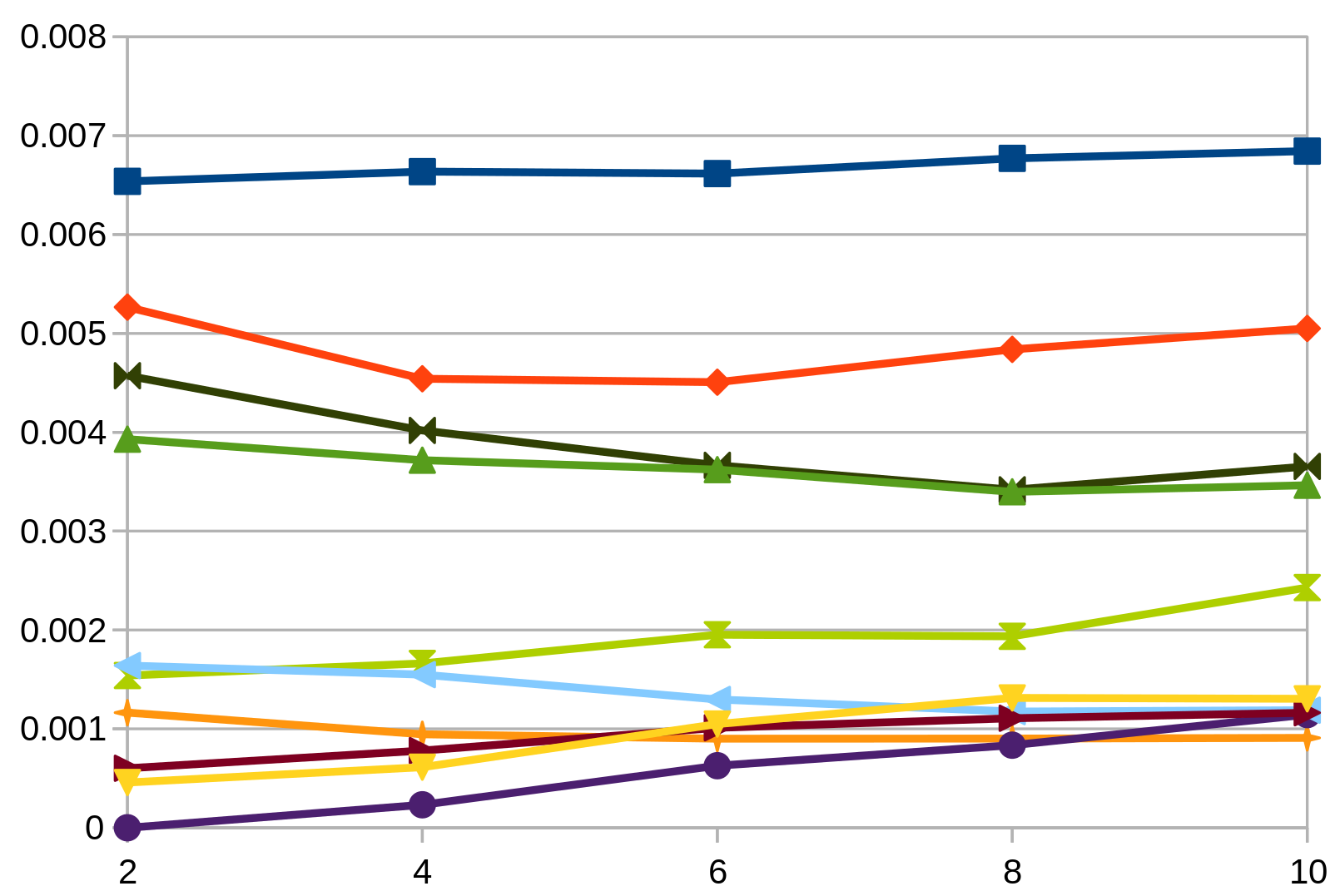}} \\ 
		%		\multicolumn{1}{{\subfloat[legend]{\includegraphics[width=2cm]{imgs/new_images/legend_small.png}}}\\
	\end{tabular}
	\subfloat{\includegraphics[width=1\textwidth]{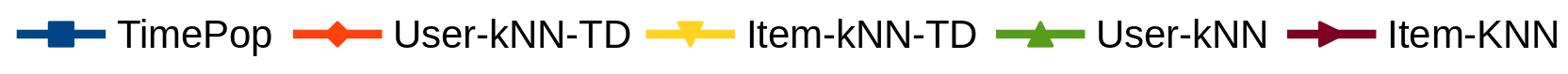}}\\ \vspace{-0.5cm}
	\subfloat{\includegraphics[width=1\textwidth]{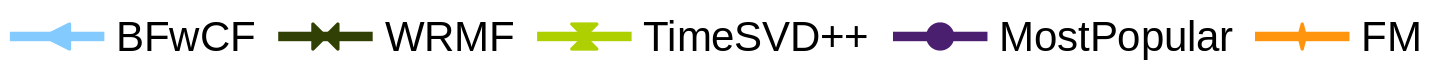}} \vspace{-0.3cm}
	\caption{nDCG @N varying N in 2..10} \vspace{-0.9cm}
	\label{fig:accuracy_results}
\end{figure*}

\paragraph{Baselines.} We evaluated our approach w.r.t  CF  and time-aware techniques. \textbf{MostPopular} was included as \framework is a time-aware variant of ``Most Popular''. From  model-based collaborative filtering approaches we selected some of the best performing matrix factorization algorithms \textbf{WRMF} trained with a regularization parameter set to 0.015, $\alpha$ set to 1 and 15 iterations, and \textbf{FM}\footnote{\url{https://github.com/sisinflab/recommenders}\label{adhoc}}\cite{rendle2010factorization}, computed with an ad-hoc implementation of a 2 degree factorization machine considering users and items as features, trained using Bayesian Personalized Ranking Criterion\cite{DBLP:conf/uai/RendleFGS09}. 
Moreover, we compared our approach against the most popular memory-based kNN algorithms, \textbf{Item-kNN}\textsuperscript{\ref{adhoc}} and \textbf{User-kNN}\textsuperscript{\ref{mym}} \cite{sarwar2000analysis}, together with their time-aware variants (\textbf{Item-kNN-TD}\textsuperscript{\ref{adhoc}}, \textbf{User-kNN-TD}\textsuperscript{\ref{adhoc}})\cite{ding2005time}. We included \textbf{TimeSVD++}\textsuperscript{\ref{adhoc}} \cite{koren2010collaborative} in our comparison even though this latter has been explicitly designed for the rating prediction task.
% while \framework computes a \TopN recommendation list. We included TimeSVD++ as it is one of the most important advances in time-aware RS. 
All model-based algorithms were trained using 10, 50, 100, and 200 factors; only best models are reported in the evaluation: for Movielens1M WRMF 10, FM 10; for Amazon Movies WRMF 100, FM 200; for Amazon Toys and Games WRMF 100, FM 50.
Finally \textbf{BFwCF} \cite{DBLP:conf/recsys/BelloginS17} is an algorithm that takes into account interaction sequences between users and it uses the last common interaction to populate the candidate items list. In this evaluation we included the BFwCF variant that takes advantage of similarity weights per user and two time windows, left-sided and right-sided (Backward-Forward).  BFwCF was trained using parameters from \cite{DBLP:conf/recsys/BelloginS17}: 100 neighbors, \textit{indexBackWards} and \textit{indexForwards} set to 5, normalization and combination realized respectively via \textit{DummyNormalizer} and \textit{SumCombiner}. Recommendations were computed with the implementation publicly provided by authors. In order to guarantee a fair evaluation, for all the time-based variants the $\beta$ coefficient was set to $\frac{1}{200}$ \cite{koren2010collaborative}. TimeSVD++ was trained using parameters used in \cite{koren2010collaborative}.
\\\textbf{Results Discussion.} Results of experimental evaluation are shown in Figure \ref{fig:accuracy_results} which illustrate nDCG (\ref{fig:accuracy_results}a, \ref{fig:accuracy_results}b, \ref{fig:accuracy_results}c) curves for increasing number of top ranked items returned to the user. Significance tests have been performed for  accuracy metrics using Student's t-test and p-values and they result consistently lower than 0.05.  By looking at Figure \ref{fig:accuracy_results}a we see that \framework outperforms comparing algorithms in terms of accuracy on AmazonMovies dataset. We also see that algorithms exploiting a Temporal decay function perform well w.r.t. their time-unaware variants (User-kNN and Item-kNN) while matrix factorization algorithms (WRMF ,TimeSVD++ and FM) perform quite bad. The low performance of MF algorithms is very likely due to the temporal splitting that makes them unable to exploit collaborative information. We may assume that the good performance of \framework w.r.t. kNN algorithms are  due to the adopted threshold, that emphasizes the popular items, and hence increases accuracy metrics values. Results for Amazon Toys and Games dataset are analogous to those computed for Amazon Movies.  Results for Movielens additionally show that the high number of very popular items make neighborhood-based approaches perform similarly.

\vspace{-0.3cm}
\section{Conclusion}\label{sec:conclusion}
%\vspace{-0.3cm}
\vspace{-0.3cm}
In this paper we presented \framework, a framework that exploits local popularity of items combined with temporal information to compute top-N recommendations.
The approach relies on the computation of a set of time-aware neighbors named Precursors that are considered the referring population for a user we want to serve recommendations. 
We compared \framework against  state-of-art algorithms showing its effectiveness in terms of accuracy despite its lower computational cost in computing personalized recommendations.


\begin{thebibliography}{10}\vspace{-0.4cm}
	\bibitem{adomavicius2001multidimensional}
	G.~Adomavicius and A.~Tuzhilin.
	\newblock Multidimensional recommender systems: a data warehousing app.roach.
	\newblock {\em Electronic commerce}, pp. 180--192, 2001.
	
	\bibitem{adomavicius2011context}
	G.~Adomavicius and A.~Tuzhilin.
	\newblock Context-aware recommender systems.
	\newblock In {\em Recommender systems handbook}, pp. 217--253, 2011.
	
	\bibitem{anelli2017feature}
	V.~Anelli, T.~Di~Noia, E.~Di~Sciascio, and P.~Lops.
	\newblock Feature factorization for top-n recommendation: From item rating to
	features relevance.
	\newblock In Proc. of RecSysKTL, pp. 16--21, 2017.
	
	\bibitem{DBLP:journals/dss/BaoLLSG13}
	H.~Bao, Q.~Li, S.~S. Liao, S.~Song, and H.~Gao.
	\newblock A new temporal and social pmf-based method to predict users'
	interests in micro-blogging.
	\newblock {\em Decision Supp.ort Systems}, 55(3):698--709, 2013.
	
	\bibitem{DBLP:conf/recsys/BelloginS17}
	A.~Bellog{\'{\i}}n and P.~S{\'{a}}nchez.
	\newblock Revisiting neighbourhood-based recommenders for temporal scenarios.
	\newblock In Proc. of TempRec, pp.  40--44, 2017.
	
	\bibitem{DBLP:journals/umuai/CamposDC14}
	P.~G. Campos, F.~D{\'{\i}}ez, and I.~Cantador.
	\newblock Time-aware recommender systems: a comprehensive survey and analysis
	of existing evaluation protocols.
	\newblock {\em UMAI}, 24(1-2):67--119, 2014.
	
	\bibitem{Cremonesi2010}
	P.~Cremonesi, Y.~Koren, and R.~Turrin.
	\newblock Performance of recommender algorithms on top-n recommendation tasks.
	\newblock In Proc. of RecSys '10, pp. 39--46, 2010.
	
	\bibitem{ding2005time}
	Y.~Ding and X.~Li.
	\newblock Time weight collaborative filtering.
	\newblock In Proc. of CIKM '05, pp. 485--492. ACM, 2005.
	
	\bibitem{fernandez2016alleviating}
	I.~Fern{\'a}ndez-Tob{\'\i}as, M.~Braunhofer, M.~Elahi, F.~Ricci, and
	I.~Cantador.
	\newblock Alleviating the new user problem in collaborative filtering by
	exploiting personality information.
	\newblock {\em UMUAI}, 26(2-3):221--255,
	2016.
	
	\bibitem{DBLP:reference/sp/GunawardanaS15}
	A.~Gunawardana and G.~Shani.
	\newblock Evaluating recommender systems.
	\newblock In {\em Recommender Systems Handbook}, pp. 265--308. 2015.
	
	\bibitem{DBLP:conf/recsys/JamborW10}
	T.~Jambor and J.~Wang.
	\newblock Optimizing multiple objectives in collaborative filtering.
	\newblock In  Proc. of RecSys '10, pp. 55--62, 2010.
	
	\bibitem{JannachLGB13}
	D.~Jannach, L.~Lerche, F.~Gedikli, and G.~Bonnin.
	\newblock What recommenders recommend - an analysis of accuracy, popularity,
	and sales diversity effects.
	\newblock In Proc. of UMAP '13, pp. 25--37, 2013.
	
	\bibitem{DBLP:journals/eswa/JugovacJL17}
	M.~Jugovac, D.~Jannach, and L.~Lerche.
	\newblock Efficient optimization of multiple recommendation quality factors
	according to individual user tendencies.
	\newblock {\em Expert Syst. App.l.}, 81:321--331, 2017.
	
	\bibitem{koren2010collaborative}
	Y.~Koren.
	\newblock Collaborative filtering with temporal dynamics.
	\newblock {\em Communications of the ACM}, 53(4):89--97, 2010.
	
	\bibitem{lathia2009temporal}
	N.~Lathia, S.~Hailes, and L.~Capra.
	\newblock Temporal collaborative filtering with adaptive neighbourhoods.
	\newblock In Proc. of SIGIR '09, pp. 796--797,
	2009.
	
	\bibitem{liu2010online}
	N.~N. Liu, M.~Zhao, E.~Xiang, and Q.~Yang.
	\newblock Online evolutionary collaborative filtering.
	\newblock In Proc. of RecSys '10, pp. 95--102, 2010.
	
	\bibitem{DBLP:conf/icdm/OhPYSP11}
	J.~Oh, S.~Park, H.~Yu, M.~Song, and S.~Park.
	\newblock Novel recommendation based on personal popularity tendency.
	\newblock In Proc. of ICDM '11, pp. 507--516, 2011.
	
	\bibitem{DBLP:conf/icdm/Rendle10}
	S.~Rendle.
	\newblock Factorization machines.
	\newblock In Proc. of ICDM '10, pp. 995--1000, 2010.
	
	\bibitem{DBLP:conf/uai/RendleFGS09}
	S.~Rendle et al.
	\newblock {BPR:} bayesian personalized ranking from implicit feedback.
	\newblock In Proc. of {UAI} '09, pp. 452--461, 2009.
	\vspace{-0.02cm}
	\bibitem{sarwar2000analysis}
	B.~Sarwar, G.~Karypis, J.~Konstan, and J.~Riedl.
	\newblock Analysis of recommendation algorithms for e-commerce.
	\newblock In Proc. of EC '00, pp. 158--167, 2000.
	\vspace{-0.02cm}
	\bibitem{steck2013evaluation}
	H.~Steck.
	\newblock Evaluation of recommendations: rating-prediction and ranking.
	\newblock In Proc. of RecSys'13, pp. 213--220, 2013.
	\vspace{-0.02cm}
	\bibitem{DBLP:conf/wsdm/WuABSJ17}
	C.~Wu, A.~Ahmed, A.~Beutel, A.~J. Smola, and H.~Jing.
	\newblock Recurrent recommender networks.
	\newblock In Proc. of {WSDM} 2017,
	pp. 495--503, 2017.
	\vspace{-0.02cm}
	\bibitem{xia2010dynamic}
	C.~Xia, X.~Jiang, S.~Liu, Z.~Luo, and Z.~Yu.
	\newblock Dynamic item-based recommendation algorithm with time decay.
	\newblock In Proc. of ICNC '10, pp. 242--247, 2010.
	\vspace{-0.02cm}
	\bibitem{zimdars2001using}
	A.~Zimdars, D.~M. Chickering, and C.~Meek.
	\newblock Using temporal data for making recommendations.
	\newblock In Proc. of UAI '01, pp. 580--588,
	2001.
	\vspace{-0.02cm}
	\bibitem{rendle2010factorization}
	S.~Rendle.
	\newblock Using temporal data for making recommendations.
	\newblock In Proc. of ICDM '10, pp. 995--1000,
	2001.
\end{thebibliography}
\end{document}